\documentclass[twocolumn,showpacs]{revtex4-1}

\usepackage{amsmath}
\usepackage{graphicx}

\renewcommand{\vec}{\boldsymbol}

\newcommand{\dd}{\mathrm{d}}

\newcommand{\includeSubGraphics}[2]{%
  \parbox{0.495\linewidth}{%
    {\raggedright#1\hfill\\[-0.33\baselineskip]}%
    \includegraphics{#2}%
  }%
}

\begin{document}

\title{Multiple collisions in turbulent flows}

\author{Michel Vo\ss kuhle$^1$, Emmanuel L\'ev\^eque$^{1,2}$, Michael Wilkinson$^3$ and Alain Pumir$^1$}

\affiliation{
 $^1$ Laboratoire de Physique, ENS de Lyon and CNRS, UMR5672,
46, all\'ee d'Italie, F-69007~Lyon, France \\
 $^2$ Laboratoire de M\'ecanique des Fluides et d'Acoustique,
Ecole Centrale de Lyon and CNRS, UMR5509, 36 avenue Guy de Collonge
F-69134, Ecully, France \\
 $^3$ Department of Mathematics and Statistics, The Open University, Walton Hall, Milton Keynes, MK6~7AA, England
}

\begin{abstract}
In turbulent suspensions, collision rates determine how rapidly particles
coalesce or react with each other.
To determine the collision rate, many numerical studies rely on the
\lq Ghost Collision Approximation' (GCA), which simply records
how often pairs of point particles come within a threshold distance.
In many applications, the suspended particles stick (or in the
case of liquid droplets, coalesce) upon collision, and it is the frequency
of first contact which is of interest. If a pair of \lq ghost' particles
undergoes multiple collisions, the GCA may overestimate the true
collision rate. Here, using fully resolved Direct Numerical
Simulations
of turbulent flows at moderate Reynolds
number (${\rm R}_\lambda = 130$), we investigate the prevalence
and properties of multiple collisions. We demonstrate that the GCA leads
to a systematic overestimate of the collision rate, which is of the order of
$\sim 15\,\%$ when the particle inertia is small, and slowly decreases when
inertia increases. We investigate the probability $P(N_{\rm c})$ for a given
pair of ghost particles colliding $N_{\rm c}$ times. We find $P(N_{\rm c})=\beta \alpha^{N_{\rm c}}$ for $N_{\rm c}>1$,
where $\alpha$ and $\beta$ are coefficients which depend upon the
particle inertia. This result is used to explain the discrepancy between the
GCA and the true collision rates. We also investigate the statistics
of the times that ghost particles remain in contact. We show that the
probability density function
of the contact time is different
for the first collision. The difference is explained by the effect
of caustics in the phase space of the suspended particles.
\end{abstract}

\maketitle

\section{Introduction}

Collisions between particles transported by a turbulent flow play a crucial
role in several important phenomena. Saffman and Turner~\cite{saffman56jfm}
suggested that
turbulence in clouds can lead to a very significant enhancement of the rate
of collision between small droplets. This mechanism has been proposed to provide
an explanation for the very fast rate of coalescence reported in warm cumulus
clouds~\cite{falkovich02nat,shaw03arfm,wilkinson06prl}. Also, collisions
between dust grains {in a turbulent circumstellar accretion disc} play an important role in some theories for planet formation~\cite{Safranov:72,wilkinson08apj}.

There is a substantial literature on the collision rate of particles in
turbulent
flows.
Most of the studies devoted to collisions in turbulent suspensions explicitly
deal with \lq geometric collisions', that is, merely detect when
the centers of two particles are separated by a distance $d$ less than the
sum of their radii, $a_1+a_2$.
The time-dependent separation $d(t)$  may cross $a_1+a_2$ repeatedly,
and every crossing from above is considered as a new collision.
In many applications, however, the colliding particles are assumed to stick,
coalesce, or react on first contact. In these cases the multiple collisions
are spurious, and the physically relevant collision rate should only count
the frequency with which $d(t)$ decreases below
$a_1+a_2$ for the first time. The \lq ghost collision approximation'
(GCA) consists in ignoring this aspect, and in following all particles in
the flow even after they underwent collisions. While this approximation is
appealing from a numerical point of view, it has been noticed to lead to
questionable estimates of the collision rate \cite{wang98pof2,chun05pof},
because it includes spurious multiple collisions \cite{gustavsson08njp}.

The first aim of this paper is to fully characterize the shortcomings of the GCA
approximation which arise from the existence of multiple collisions.
To this end, we write the collision rate
$\Gamma$ as a sum of two terms:
\begin{equation}
\Gamma_{\rm GCA} =  \Gamma_1 + \Gamma_{\rm m}
\label{eq:gamma_decomp}
\end{equation}
where $\Gamma_{\rm GCA}$ is the collision rate obtained with the GCA.
The term $\Gamma_1$ corresponds to the collision rate for particles
undergoing a first contact, and $\Gamma_{\rm m}$ is the rate for multiple
collisions between a pair of particles that has already collided once.
Thus $\Gamma_{\rm m}$ represents the unphysical contribution to the GCA
collision rate due to multiple encounters between a given pair of particles.
One way to evaluate the true collision rate $\Gamma_1$ is to count
the rate of multiple collisions, and subtract it from $\Gamma_{\rm GCA}$.

What specifically happens to two particles after they have come into
contact depends on the precise physical
problem~\cite{Bec:12,Krstolovic:13}. We are interested here in problems
where particles have a strong probability of reacting after their first
contact. For the sake
of the present work, we assume here that $\Gamma_1$, defined by considering
only the first collisions occurring between a pair of particles, is
an appropriate definition of the collision rate. We have also considered
two other processes, which lead to alternative definitions.
The first process consists in assuming explicitly that
the two particles do not participate any longer to the reactions as soon as
they have come into contact, as if they had annihilated.
Technically, in order to deal
with a steady-state system, we introduce two new particles to compensate for
the loss. The implementation of this
algorithm requires some care in the case of inertial particles, as the
velocity depends on history. Another process consists in removing and
replacing only one of the two colliding particles.
We demonstrate here that the collision rates describing these two processes
are equal to $\Gamma_1$, in the limit of very dilute suspensions, which
justifies its relevance.

The other objective of our study is to characterize the statistics of the
multiple collisions. We investigate the statistics of the number of contacts,
$N_{\rm c}$.
We find that there is a very simple distribution of the number
of times a pair of particles collides:
the probability
for observing $N_{\rm c}$ collisions after one initial collision is found to be well approximated by
\begin{equation}
\label{eq:poisson}
P(N_{\rm c} \vert N_{\rm c} \geq 1) = \beta\,\alpha^{N_{\rm c}}
\end{equation}
where the coefficients $\alpha$ and $\beta$ depend upon parameters
as discussed below. We use these results to explain the spurious collision
rate, $\Gamma_{\rm m}$.

It has been observed that tracer particles in a turbulent flow can remain
in proximity for a long time~\cite{jullien99prl,rast11prl,scatama:13}.
This phenomenon
may be related to the multiple collisions which we are investigating here, and
could be of interest to model classes of reaction occurring with a finite
probability when particles are close enough~\cite{Krstolovic:13}.
With this motivation, we consider the statistics of the time that a
pair of ghost particles is
in contact and relate this to the distribution of relative velocities of
collisions. The probability distribution function (PDF) of
the contact time exhibits a striking structure.
For the first collision, it follows a power-law at intermediate values, and an exponential decay at longer contact times.
The PDF of the contact
time for the first collision are
markedly different from those of subsequent collisions (which lack the
power-law behavior and which appear to be independent of the number $N_{\rm c}$ of collisions, for $N_{\rm c} \ge 2$).
We explain this discrepancy by appealing to a model for the collision process
introduced in \cite{falkovich02nat} and \cite{wilkinson06prl}, according
to which the collision rate $\Gamma$ can be expressed as the sum of two
terms, resulting from two different types of physical processes:
\begin{equation}
\label{eq:advcaust}
\Gamma= \Gamma_{\rm adv}+\Gamma_{\rm caust}\ .
\end{equation}
The term $\Gamma_{\rm adv}$ represents the rate of collisions of particles
which are advected into contact by shearing motion due to turbulence.
This advective process allows
for the possibility of multiple collisions, because the local velocity
gradient
can fluctuate as a function of time, so that particles may be brought
back into contact after separating for a while.
The term $\Gamma_{\rm caust}$ results from
the possibility that particles with significant inertia can move relative to
the fluid.
If caustics (fold lines) form in the phase-space of the suspended
particles, particles occupying the same position may have different
velocities.
This phenomenon can also be understood as resulting from particles being
centrifuged out of vortices, and has also been referred to as the \lq sling'
effect~\cite{falkovich02nat,falkovich07jas}.
Experiments have recently demonstrated the existence of the
\lq sling' effect in a well-controlled laboratory flow~\cite{bewley:2013}.
Particles colliding at higher relative velocity are unlikely to undergo
multiple collisions. The higher-velocity collisions generated by caustics
contribute to the statistics of the first collision, but not to the multiple
collisions. The use of the decomposition
\eqref{eq:advcaust} is supported by work demonstrating that it
is valid for a random flow model~\cite{gustavsson11pre,gustavsson13_ar2}.
A companion to this paper \cite{vosskuhle13} shows that (\ref{eq:advcaust})
is a good model for DNS studies of turbulence, and discusses the relative
importance of the two terms as a function of the parameters of the
turbulent suspension.

We will show that the power-law PDF of contact
times is entirely due to the first
collision between particles, and not to the subsequent ones.
The subsequent collisions appear to be self-similar, in the sense that the distributions of
times when particles are within a given distance remain unchanged, no matter
how many times the particles have come close together.
We also consider possible explanations for this observation.

The results presented here very significantly extend a previous study,
using a simpler model of turbulent flow, namely the Kinematic Simulation
approach~\cite{vosskuhle11jpcs}.
While the results concerning the relative errors made by using this simplified
flow are qualitatively similar
to the results presented in this work, we find that Kinematic Simulations
lead to a very significant underestimation of the collision rates, by almost
an order of magnitude.

Our work emphasizes models where the suspended particles
coalesce upon contact, so that multiple collisions
are a source of error in the collision rate.
These models are just a limiting case of a larger class of models, which
may also be of physical interest. As an example,
the case of fully elastic collisions
was recently considered by \cite{Bec:12}, who showed that elastic collisions
lead to a finite probability of particles making an infinite number of
collisions. Our investigations of the contact times in the ghost collision
approximation involve evaluating the distribution of the time
that one particle in a turbulent flow spends within a sphere surrounding
another particle. We remark that J{\o}rgensen {\em et al} \cite{Ott:05} have investigated this
distribution experimentally in a different context, where the
radius of the sphere is much larger than the Kolmogorov length
of the flow.

This article is organized as follows.
In Section~\ref{sec:numerics}, we present
the numerical schemes used to simulate the carrying turbulent flow and account for the dynamics and collisions of the suspended particles.
We describe the algorithms used to study the physical problem of particles
coagulating during their first contact, and to establish that the reaction rate
reduces, in the very limit, to $\Gamma_1$. The nature of the correction due to
a finite particle density are discussed in depth in an Appendix.
The quantitative estimates of the errors made by using the GCA are discussed in Section \ref{sec:gca_error}, together with the data justifying the empirical
law (\ref{eq:poisson}) for a pair of particles to undergo multiple collisions.
Section~\ref{sec:contactstats} is devoted to the statistics of the
multiple collisions. Section \ref{sec:vel_diff} discusses the explanation for some of the
observations in terms of statistics of the relative velocities upon collision.
Finally, we summarize our results in Section~\ref{sec:concl}.

\section{Numerical methods}
\label{sec:numerics}

\subsection{Direct Numerical Simulation of Navier-Stokes Turbulence}

The work rests on simulating the (incompressible) Navier--Stokes equations:
\begin{eqnarray}
\partial_t \vec{u}( \vec{x}, t) + ( \vec{u}(\vec{x} ,t ) \cdot \nabla ) \vec{u}
(\vec{x} ,t ) & = &- \nabla p (\vec{x}, t)+ \nu \nabla^2 \vec{u}(\vec{x}, t)\nonumber
\\
&&+ \vec{f} ( \vec{x} , t) \label{eq:NS} \\
\nabla \cdot \vec{u} (\vec{x} , t) & = & 0 \label{eq:incomp}
\end{eqnarray}
where $\vec{u}(\vec{x} ,t )$ denotes the Eulerian velocity field, $\nu$ is the viscosity,
and $\vec{f}(\vec{x} ,t )$ is a forcing term; the mass density is arbitrarily set to unity.
These equations are solved in a cubic box of size $2\pi$ with periodic
boundary conditions in the three directions by a pseudo-spectral method.
The pressure $p (\vec{x}, t)$ is eliminated by taking the divergence of (\ref{eq:NS}) and by solving the resulting Poisson equation in the spectral domain \cite{Frisch:1995}.
The forcing term acts on Fourier modes of low wavenumbers, $| \vec{k} | \le K_f$.
It is adjusted in such a way that the injection rate of energy, $\varepsilon$, remains constant~\cite{Lamorgese:05}:
\begin{equation}
\vec{f}_{\vec{k}} = \varepsilon ~ \frac{ \vec{u}_{\vec{k}} }{\sum_{|\vec{k}| \le K_f}| \vec{u}_{\vec{k}} |^2 }
\quad \mathrm{if} \quad | \vec{k} | \le K_f
\ .
\label{eq:forcing}
\end{equation}
The simulations discussed in this study have
been done with the following parameters, in code units:
$\varepsilon = 10^{-3}$, and $\nu = 4 \times 10^{-4}$. The forcing operates at wavenumbers $| \vec{k} | \le K_f=1.5$.
With these
values, the Kolmogorov scale, $\eta_{\rm K} = (\nu^3/\varepsilon)^{1/4}$,
is comparable to the \emph{effective} spatial resolution
$\Delta x = 2 \pi/256$: $\Delta x/ \eta_{\rm K} \approx 1.5$ or $k_\mathrm{max} \eta \approx 2.1$, which fulfills standard requirements for the direct numerical simulation (DNS) of Navier-Stokes turbulence and the integration of particle trajectories \cite{calzavarini09jfm}.
Let us notice that, in practice, the number of grid points is 384 in each direction according to the two-thirds rule \cite{orszag71jas} (used to avoid aliasing errors).
However, padded high-wavenumber modes are not excited and, therefore, do not contribute to improving the spatial resolution of the solution.
With these values of the parameters, the value of the Reynolds number based on the Taylor microscale is ${\rm R}_\lambda \approx 130 $ in the statistically stationary long-time limit state.

After spectral truncation, \eqref{eq:NS} reduces to a set of ordinary differential
equations (in time) for the Fourier modes, which have been integrated
using the second-order Adams-Bashforth scheme.
The time step $\delta t$ has been chosen so that the Courant number
$\mathrm{Co}= u_\mathrm{rms} \cdot k_\mathrm{max} \delta t \lesssim 0.1$.
With this choice, the relaxation time $\tau_\mathrm{p}$ of the particle dynamics remains of the order of $10^2\cdot \delta t$, which ensures that particle trajectories can be safely integrated from the time-evolving Eulerian velocity field.

\subsection{Dynamics of particles}

We simulated in our flow the motion of small particles, all of which are
assumed to have the same size and mass, whose motion obeys
the following set of equations:
\begin{align}
 \frac{{\dd}\vec{x}}{{\dd}t} &= \vec{v}, & \frac{{\dd}\vec{v}}{{\dd}t} &= \frac{\vec{u}(\vec{x},t) - \vec{v}}{\tau_{\rm p}}.
 \label{eq:max_ril}
\end{align}
This set of equations is a simplified version of the original set derived by
\cite{maxey83pf, gatignol83jmta}, and is appropriate for small spherical
particles with radius $a$ which is much smaller than the Kolmogorov scale
$\eta_{\rm K}$, and
whose density is much larger than the fluid density:
$\rho_{\rm p}/\rho_{\rm f} \gg 1$.
In order to isolate the role of turbulence, we have
explicitly neglected gravity in \eqref{eq:max_ril}, despite the
fact that it plays an important role in cloud microphysics, sandstorms and other
terrestrial phenomena. Gravitational effects on collision rates are unimportant in applications to planet formation.
The relaxation time in \eqref{eq:max_ril} is determined by the Stokes
drag:
\begin{equation}
\tau_{\rm p} = \frac{2}{9} \frac{\rho_{\rm p}}{\rho_{\rm f}} \frac{a^2}{\nu}\ .
\end{equation}
This relaxation time
is made dimensionless by using the Kolmogorov time scale,
$\tau_{\rm K} = (\nu/\varepsilon)^{1/2}$, and the Stokes number
\begin{equation}
\mathrm{St} = \frac{\tau_{\rm p}}{\tau_{\rm K}}
\end{equation}
is a dimensionless measure of the importance of inertial effects in determining
the trajectories of the particles. In order to explore parameter ranges
which are relevant to cloud microphysics, we used a ratio of densities $\rho_{\rm p}/\rho_{\rm f} = 10^3$
throughout. Lengths and times are given in dimensionless units and can be
readily scaled to realistic situations.

The determination of the particle velocity $\vec{v}$ requires,
according to \eqref{eq:max_ril}, the
evaluation of the fluid velocity $\vec{u}$ at the location of the particle.
This is done by resorting to tri-cubic interpolation.
The particle trajectories have been integrated by using the second-order Verlet velocity algorithm \cite{press07}.

As we are interested in determining the collision rates in turbulent flows, we
simulated a large number of particles. The number of particles $N_{\rm p}$
used to monitor the collision rates was chosen
in such a way that the
volume fraction
$\Phi = 4 N_{\rm p} \pi a^3/(3L^3)$, where $L$ is the size of the system
($L = 2 \pi$), is either $\Phi = 4.5 \times 10^{-6}$ or
$\Phi = 4.5 \times 10^{-5}$. In both cases, collisions involving three or
more particles can be neglected.

In a flow having reached a statistically steady state,
we inserted at a time $T = 0$ a total number $N'_{\rm p}$ of particles, initially
distributed uniformly in the flow. We then integrated the equation of motions
\eqref{eq:NS},\eqref{eq:max_ril}, for a time of the order of $10$ eddy-turnover times, $T_L$, defined by:
\begin{equation}
T_L =  \frac{L}{\sqrt{\langle {\bf u}^2/3 \rangle}}, ~~ {\rm with} ~~ L = \frac{3 \pi}{2 \langle {\bf u}^2 \rangle} \int k^{-1} E(k) \,\dd k\ .
\label{eq:T_L}
\end{equation}
Past this time, we integrated the equations of motion for a time
$t_{tot}$ larger than $ 15 T_L$ (except for ${\rm St} = 0.2$).
All the particle trajectories were saved, with a sampling time of $\Delta t \approx  0.055 \tau_{\rm K}$, and processed afterward.
Table~\ref{tabl:1} summarizes our runs.

\begin{table*}
 \caption{(Color online)
Summary of parameters from our different DNS runs.
We tabulate the Stokes number $\mathrm{St}$, the volume fraction occupied
by the particles $\Phi$, the total integration time $t_\text{tot}$
used to determine the collision rate, expressed in
terms of the large eddy turnover time $T_L$,
and the total number of collisions $N_{\rm GCA}$ detected, when using the ghost collision approximation.
}
 \label{tabl:1}
 \begin{tabular}{lcccccccccccccc}
  \hline\hline
  $\mathrm{St}$ & $0.0$ & $0.10$ & $0.20$ & $0.30$ & $0.51$ & $0.76$ & $1.01$ & $1.27$ & $1.52$ & $2.03$ & $2.53$ & $3.04$ & $4.05$ & $5.07$\\
  \hline
  $\Phi \times 10^{6}$ & $4.5$ & $4.5$ & $45$ & $4.5$ & $45$ & $45$ & $45$ & $45$ & $45$ & $45$ & $45$ & $45$ & $45$ & $45$\\
  $t_\text{tot} / T_L$ & $15.5$ & $15.5$ & $4.2$ & $31.4$ & $15.9$ & $10.4$ & $47.6$ & $13.0$ & $52.4$ & $52.1$ & $52.5$ & $52.3$ & $53.1$ & $42.1$\\
  $N_{\rm GCA} / 10^{4}$ & $0.6$ & $1.3$ & $29$ & $3.3$ & $200$ & $180$ & $400$ & $81$ & $250$ & $150$ & $99$ & $72$ & $42$ & $22$\\
  \hline\hline
 \end{tabular}
\end{table*}

For the case of $\mathrm{St} = 0$, we integrated the trajectories of Lagrangian tracer particles.
Those are point particles with no extent, but to determine the collision rate it is necessary to assume they have a finite size.
We chose the radius to be the same as for particles with $\mathrm{St} = 0.1$.

\subsection{Detection of the collisions}
\label{sec:detect}

In a monodisperse solution of volume $V$, containing $N_{\rm p}$ particles, the
average number of collisions ${\cal N}_{\rm c}$ per unit of time and per unit volume,
is proportional to the square of the average number of particles per unit
volume, $n \equiv N_{\rm p}/V$:
\begin{equation}
 {\cal N}_{\rm c} = \Gamma_{\rm c} \frac{n^2}{2}.
\label{eq:def_gam}
\end{equation}
Eq.\,\eqref{eq:def_gam}
defines the collision kernel $\Gamma_{\rm c}$, which depends
both on the fluid motion, and on the physical
properties of the particles.

The particle trajectories determined numerically were stored and post-processed
separately, to determine the number of collisions and other statistics.
Detecting collisions requires checking the mutual distance between $N_{\rm p}$
particles, which typically requires of the order of $N_{\rm p}^2$ operations. In
simulations such as ours involving a large number of particles, this can
lead to a prohibitively large computational time.
We used an algorithm based on a cell-linked list to speed-up the processing
time, as done in~\cite{sundaram96jcp}.

Each data set was processed in two different ways:
\begin{enumerate}
\item

{\sl Multiple collisions of ghost particles}

First we determined the collision rate using the ghost collision approximation.
In this case only $N_{\rm p}$ of the total $N'_{\rm p}$ simulated particles were used.
The collision kernel $\Gamma_{\rm GCA}$ was simply determined by counting the total
number of detected collisions ${\bf N}_{\rm c}$.
In addition, we also determined the rates
at which pairs of particles come into contact for the ${N_{\rm c}}^{\rm th}$ time, described by the collision kernel $\Gamma_{N_{\rm c}}$.
To this end, at each collision between two particles, we examine the
trajectories leading up to the collision
and determine the number of previous collisions between the same
pair. If the pair has undergone $N_{\rm c}-1$ previous collisions with each other, the collision  event
contributes to the kernel $\Gamma_{N_{\rm c}}$. Because these collision kernels describe
an exhaustive and mutually exclusive decomposition of
$\Gamma_{\rm GCA}$, we have
\begin{equation}
\label{eq:Gamsum}
\Gamma_{\rm GCA} = \sum_{N_{\rm c}=1}^\infty \Gamma_{N_{\rm c}}
\end{equation}
The collision kernel of multiple collisions $\Gamma_{\rm m}$ is simply defined by
summing the kernels $\Gamma_{N_{\rm c}}$, for $N_{\rm c} \ge 2$:
\begin{equation}
\label{eq:Gammult}
\Gamma_{\rm m}= \sum_{N_{\rm c}=2}^\infty \Gamma_{N_{\rm c}}
\ .
\end{equation}

\item

{\sl Collision detection with particle replacement.}

We have assumed that, in the case of particles coalescing or reacting on their first contact, $\Gamma_1$ is the correct measure of the collision rate.
To establish this result, we compared $\Gamma_1$ with a collision rate
algorithm where particles are removed from the flow after collision.
%
%
We determined the collision rate
again with $N_{\rm p}$ particles, but we systematically
replaced either one or two of the colliding particles after each collision.
The substitutions are carried out by simply picking one of the $(N'_{\rm p} - N_{\rm p})$
non-colliding particles, making sure that the newly introduced particle
is not colliding with any other particle at the moment of its insertion.
In so doing, the collision rate is determined at a fixed density.
A similar method has been used in Refs.\,\cite{wang98pof2,zhou98pof}.
The rates determined by this procedure are denoted $\Gamma_{{\rm Re}1}$
and $\Gamma_{{\rm Re}2}$, depending upon whether one or two
particles are replaced.
We stress that the simplest method, consisting in generating new
particles at random positions,
cannot work in the case of finite inertia (${\rm St} \ne 0$), as the velocity
depends on history.
Simulating $N'_{\rm p} > N_{\rm p}$ particles allows us to deal only with particles which
are already in equilibrium with the flow. We refer in the following to
these algorithms as the one-
or two-particle \lq substitution schemes'.

It is not immediately clear that
$\Gamma_{{\rm Re}1}=\Gamma_{{\rm Re}2}=\Gamma_1$.
In fact, our numerical results reveal measurable deviations from these
identities at the highest particle volume fraction studied here. However, as
we explain in detail in the Appendix, these deviations decrease linearly
when reducing the particle density, thus establishing that $\Gamma_1 =
\Gamma_{{\rm Re}i}$ in the limit of very dilute suspensions.

\end{enumerate}

In all cases the number of collisions ${\bf N}_{\rm c}(\Delta\tau)$ is measured in consecutive intervals of length $\Delta \tau \sim T_L$.
For each interval the collision kernel can be determined as
\begin{equation}
\langle \Gamma \rangle_{\Delta \tau}
= \frac{2 V {\bf N}_{\rm c}(\Delta\tau)}{\Delta\tau\, N_{\rm p}^2},
\label{eq:coll_rate}
\end{equation}
which is simply the collision rate divided by $n^2 / 2$.
These \lq instantaneous' collision kernels vary in time.
Determining the level of fluctuations of the number of collisions
recorded over a limited time interval led to an estimate of the
uncertainty of the collision rate.
The resulting uncertainty in the numerial value of $\Gamma_{\rm GCA}$ is
less than $2\, \%$, except for $\mathrm{St} = 0.2$ and $\mathrm{St} = 0.5$,
where it is no larger than $4\, \%$.

\section{Quantifying the error induced by the ghost collision approximation}
\label{sec:gca_error}

\subsection{Error of the Ghost-Collision Approximation}
\label{subsec:gca}

\begin{figure*}
 \includeSubGraphics{(a)}{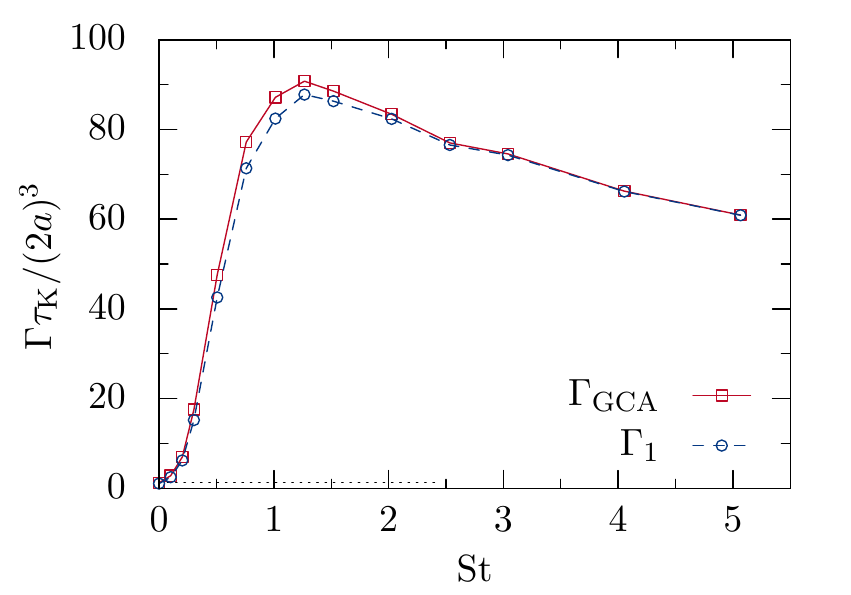}
 \includeSubGraphics{(b)}{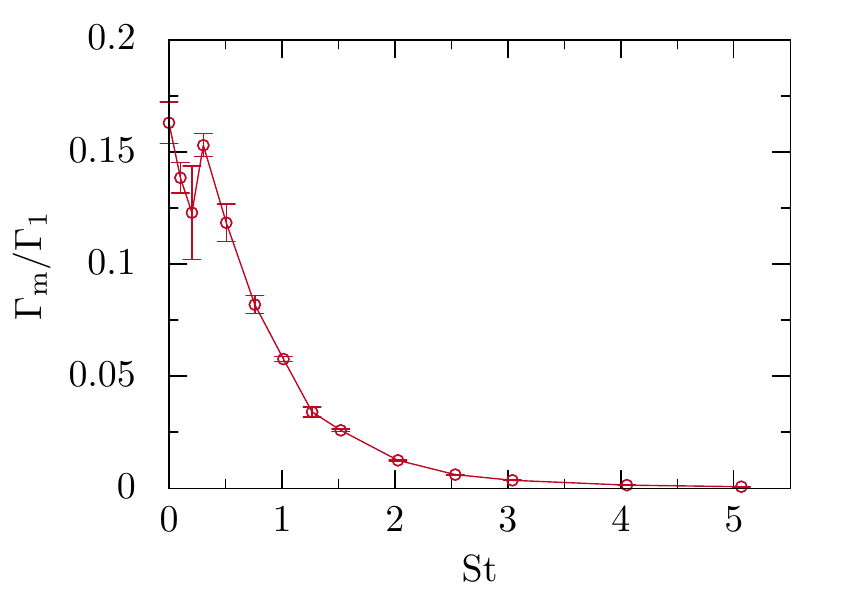}
\caption{(Color online)
Comparison between the collision kernels $\Gamma_{\rm GCA}$ and $\Gamma_1$.
The collision kernel $\Gamma_{\rm GCA}$ in panel (a) was obtained according to the ghost collision approximation, taking into account all collisions.
The collision kernel $\Gamma_1$ is restricted to only first collisions of a same pair.
The difference, $\Gamma_{\rm m} = \Gamma_{\rm GCA} - \Gamma_1$, which
corresponds to the relative error introduced by the GCA, is shown in the left panel.
This error starts at a value close to $\sim 15\%$ when ${\rm St} \rightarrow 0$ and decreases for higher Stokes numbers.
}
\label{fig:error_gca}
\end{figure*}

In this section, we discuss the error induced by using the ghost collision
approximation.
The  left
panel of Fig.~\ref{fig:error_gca} shows the collision rate computed by using
the ghost collision approximation (curve with square symbols),
and the collision rate $\Gamma_1$ (curve with circle symbols).
Our estimates of the collision rates compare quantitatively very
well with previous numerical work~\cite{sundaram97jfm,rosa:13}.
The  right panel shows the relative difference
$\Gamma_{\rm m} / \Gamma_{1} = (\Gamma_{\rm GCA}-\Gamma_{1}) / \Gamma_{1}$.

Our results show that among all the collisions recorded, as many as $15\%$
of them involve pairs of particles that have already come into
contact before.
We note that our results show that the collision rate determined by
using the GCA is correctly approximated by the Saffman--Turner formula when
${\rm St } \rightarrow 0$, consistent with previous results~\cite{wang98pof2}.
However, in the limit of small inertia, the property that pairs of particles
collide more than one time is very significant. This probability decreases
when the Stokes number increases.
This result can be qualitatively
understood by using the known fact that when the Stokes number increases,
the particle trajectories differ more from the fluid trajectories, which allows
collisions between particles with an increasingly large velocity difference.
Colliding particles are therefore expected to separate quicker when
the Stokes number is large, thus making multiple collisions less likely.

In physical situations where particle pairs react upon first
contact, the collision rate $\Gamma_1$ is expected to provide the correct
estimate of the reaction rate. To this end, we have measured the
rates $\Gamma_{{\rm Re}_1}$ and $\Gamma_{{\rm Re}_2}$, defined by taking one or two
of the colliding particles out of the system after collisions. The values
of $\Gamma_{{\rm Re}_1}$ and $\Gamma_{{\rm Re}_2}$ are found to be actually
{\it smaller} than $\Gamma_1$. In fact, as we explain in the Appendix, the
difference between $\Gamma_{{\rm Re}_i}$ ($i = 1$, $2$) and $\Gamma_1$ is due
to close encounters between three particles, an effect whose relative importance
becomes weaker when the total density of particles goes to zero. The
analysis presented in the Appendix establishes that in the limit of a very
dilute system, the proper collision rate is indeed $\Gamma_1$.

\subsection{Statistics of multiple collision }
\label{subsec:multiple}

\begin{figure*}
\includeSubGraphics{(a)}{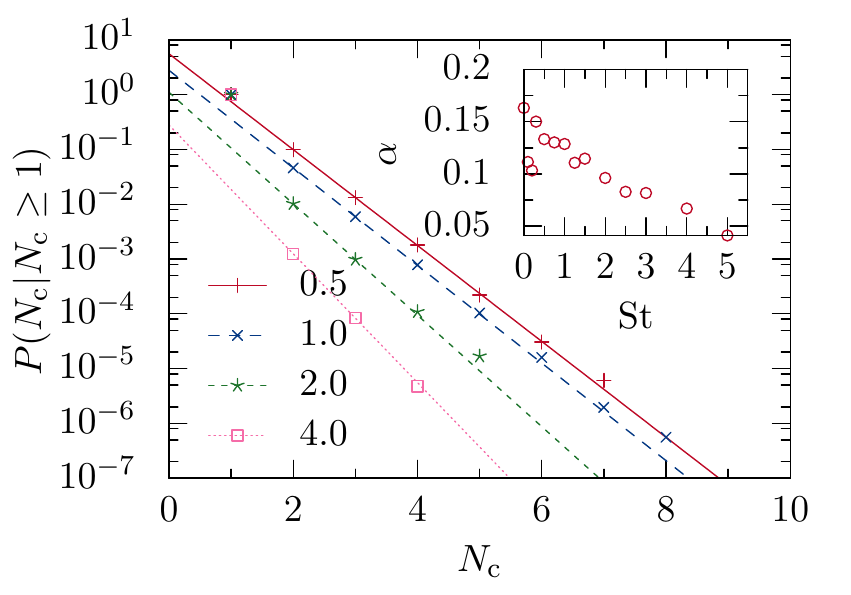}
\includeSubGraphics{(b)}{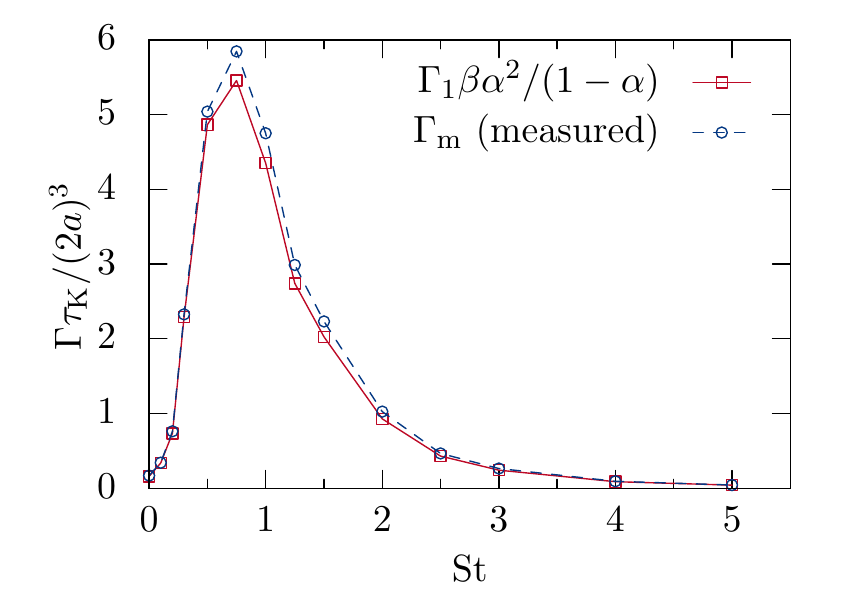}
\caption{(Color online)
Panel~(a):
The probability that a pair of particles collides an ${N_{\rm c}}^{\rm th}$ time conditioned on the fact, that it collides at least once.
The probability of observing $N_{\rm c}$ collisions goes as $\beta \alpha({\rm St})^{N_{\rm c}}$.
The results are shown at different Stokes numbers. The values
of $\alpha({\rm St})$ are shown in the inset of the figure as a function of
${\rm St}$.
Panel~(b):
The kernel for multiple collisions $\Gamma_m$ determined from
Eq.\,\eqref{eq:errsum} with the fitting parameters deduced from panel
(a) square
symbols), and measured directly in our simulations (circles).
The quantitative agreement between both results confirms the consistency
of our reasoning and the quality of our fits.
}
\label{fig:multiple_coll}
\end{figure*}

In view of the importance of multiple collisions between a pair
of particles, we characterize here the statistical properties of the number
of collisions a given pair of particles undergoes before separating.
Fig.~\ref{fig:multiple_coll} shows the probability distribution function for
a particle pair to collide an ${N_{\rm c}}^{\rm th}$ time after at least one initial collision.
The clear result from Fig.~\ref{fig:multiple_coll} is that for $N_{\rm c} \ge 2$, the
probability distribution is very well approximated by an exponential law
of the form of equation (\ref{eq:poisson}).
This remarkably simple functional form leads to the interpretation that
once a
pair of particles has undergone more than one collision, it has a probability $\alpha$ to collide once more before it separates,
the quantity
$\alpha$ being independent of $N_{\rm c}$. This  suggests
a Markovian process of multiple collisions, amenable to a simple modeling.

The exponential law, equation (\ref{eq:poisson}), can be used to
sum the series in equation (\ref{eq:Gamsum}) and hence to determine the error
of the $\Gamma_{\rm GCA}$ estimate. We have
$\Gamma_{N_{\rm c}}=\Gamma_1 \beta \alpha^{N_{\rm c}}$ for
$N_{\rm c} > 1$, so that
\begin{equation}
\label{eq:errsum}
\Gamma_{\rm m} = \Gamma_1 \beta \sum_{N_{\rm c} =2}^\infty \alpha^{N_{\rm c}}
 = \Gamma_1 \frac{\alpha^2 \beta}{1 - \alpha}\ .
\end{equation}
The numerical values obtained from expression \eqref{eq:errsum}, with
the values of $\alpha$ and $\beta$ extracted from the probability of
$N_c$ (see Fig.~\ref{fig:multiple_coll}(a))
agrees quantitatively very well with the value of $\Gamma_m$ determined
directly
(see Fig.~\ref{fig:multiple_coll}(b)).
(the two values differ by less than $10\, \%$).

\section{Statistics of the contact time between particle pairs}
\label{sec:contactstats}

Our numerical observation that a given pair of particles may collide many
times in a turbulent flow can be related to some surprising properties of
particle trajectories, which have been partly documented
before~\cite{jullien99prl,rast11prl,scatama:13}
mostly in the case of tracers (${\rm St} = 0$). In
this section, we fully characterize several properties concerning the time
particle pairs spend together, which are relevant to the subject of the
present article.

\subsection{Particle trajectories can stay close together for a long time}
\label{subsec:obs}

\begin{figure*}
\includeSubGraphics{(a)}{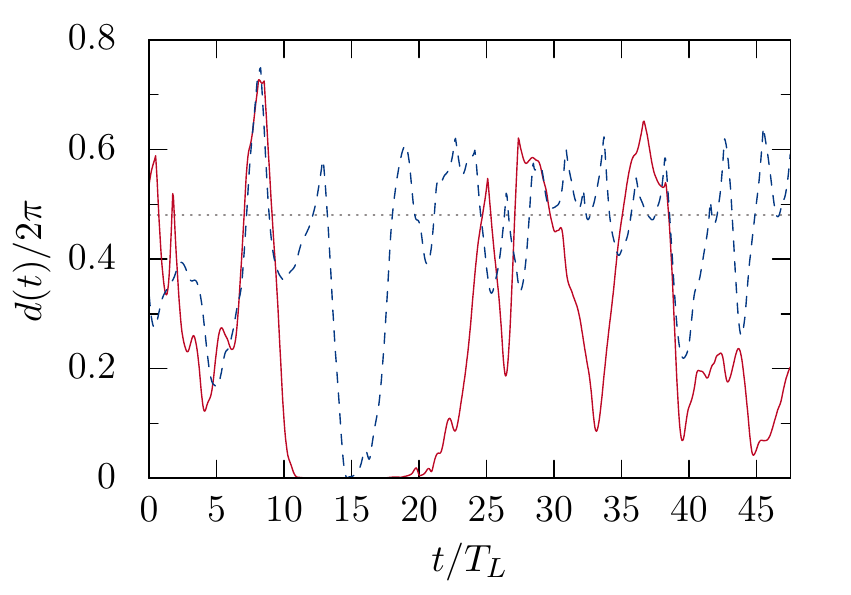}
\includeSubGraphics{(b)}{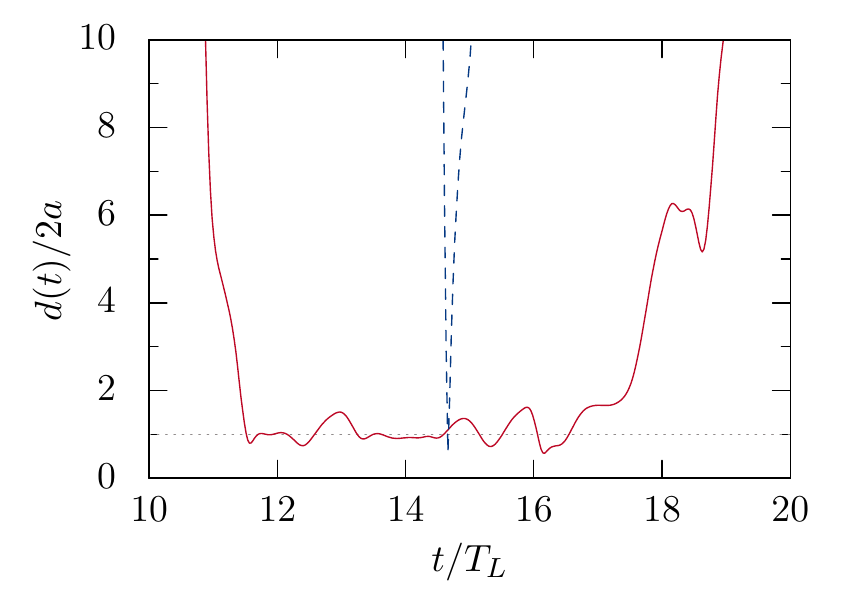}
\caption{(Color online)
Distance between two different pairs of particles as a function of
time. Panel (a) shows the distance over several eddy turnover times, and shows
that particles can be close to each other for a short time (dashed curve), or
for a much longer time (full line), corresponding to
several large eddy turnover times $T_L$. Panel (b) magnifies the graph
shown in panel (a) in the region where the distance between the particles is
very small. The full line reveals that multiple collisions occur.
The dotted line in panel (a) is the average distance for particle pairs homogeneously distributed in a periodic cube.
In panel (b), the dotted line indicates the collision radius $2a$.
}
\label{fig:close_traj}
\end{figure*}

Fig.~\ref{fig:close_traj} shows the distance between
two pairs of particles over a long time. Panel (a) shows
the distance in units of the size of the periodic box during one entire simulation.
As expected, the particles are far from each other for most of the
time.
However, we notice that the distance can reach a very small value at some particular
moments. Whereas one of the pairs of particles (dashed line)
is close to each other only for a short while, the other pair spends a much longer time close to each other.
Panel (b) blows-up the result in the range of time where the distance is small.
(Here the distance is shown in terms of the collision radius $2a$.)
The pair whose distance is plotted with a dashed line reaches the value of $2a$,
{\it i.e.} collides, but bounces apart immediately. In comparison, the
continuous curve shows that the two particles stay together for a time larger
than the large eddy-turnover time. Over this period of time, the distance
fluctuates close to the value of $(2a)$, which causes multiple collisions
between the two particles. The value of the Stokes number of the particles
shown in Fig.~\ref{fig:close_traj} is ${\rm St} = 1.0$; the phenomenon
shown here is qualitatively similar at different Stokes numbers.

We observe that the timescale for these multiple collision processes is
comparable to
the turnover time of the largest eddies in our simulation $T_L$. In
the following
investigations, we have expressed our results in terms of a dimensionless
time $t/T_L$. However the ratio between $T_L$ and the Kolmogorov
time scale, $\tau_{\rm K}$ in our
simulations is not very large: $T_L/\tau_{\rm K}\approx 10.5$.
At present it is
unclear whether $t/T_L$ is the most natural dimensionless
timescale. We remark that a plausible alternative is $t \, \lambda_1$,
where $\lambda_1$ is the leading Lyapunov exponent. Because
 $\lambda_1 \approx 0.15/\tau_{\rm K}$
\cite{bec06pof}, the variables $t/T_L$ and $t \, \lambda_1$ are
quite similar in magnitude.

\subsection{Distribution of contact times}
\label{subsec:Stat_traj}

\begin{figure}
\includegraphics{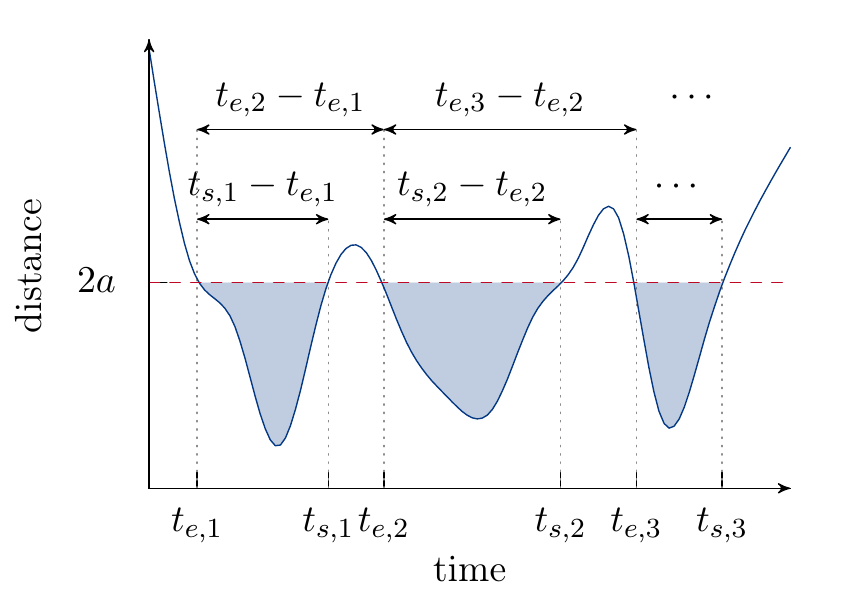}
\caption{(Color online)
Illustration of the definition of the different times: time of first encounter $t_{e,1}$, time of first separation $t_{s,1}$, time of second
encounter $t_{e,2}$ and of second separation $t_{s,2}$.
\label{fig:times_illustration}}
\end{figure}

The phenomenon reported here is very reminiscent of the observation discussed
in several earlier works~\cite{jullien99prl,rast11prl,scatama:13},
namely that particles can stay close for a very long time. While this
property has been documented mostly in the case of tracers, our observations
suggest that inertial particles can also remain very close for
a long time.
As it has been noticed in a slightly different context~\cite{Krstolovic:13},
it is of general interest to characterize the statistical properties of
the time that particles spend together.

In this subsection, we determine quantitatively the statistical properties
of the time particle trajectories remain close to one another.
We illustrate in the two following subsections the statistics at
a fixed value ${\rm St} = 1.5$ of the Stokes number.

Fig.~\ref{fig:times_illustration} introduces our notation. We consider
a pair of particles which become closer than a threshold distance $d_c$ at
an instant of time. We denote by $t_{e,1}$ the first instant for which the
distance between the particles becomes less than $d_c$,
(that is, the instant of their first collision)
and then, $t_{s,1}$
the first instant $t_{s,1} > t_{e,1}$ when they separate.
If particles
approach to within a distance $d_c$ again at some later time, we denote by
$t_{e,2}$ the first instant for which the distance becomes again smaller than
$d_c$, and $t_{s,2}$ the time at which the particles separate again. This
notation can be easily generalized when the trajectories become more than two
times closer than $d_c$.
The time particles spend together during their first encounter is denoted
$\Delta T_1 \equiv (t_{s,1} - t_{e,1})$, which again, can be easily generalized to the time particles spend together during their $n^{\rm th}$ encounter.

\begin{figure*}
\includeSubGraphics{(a)}{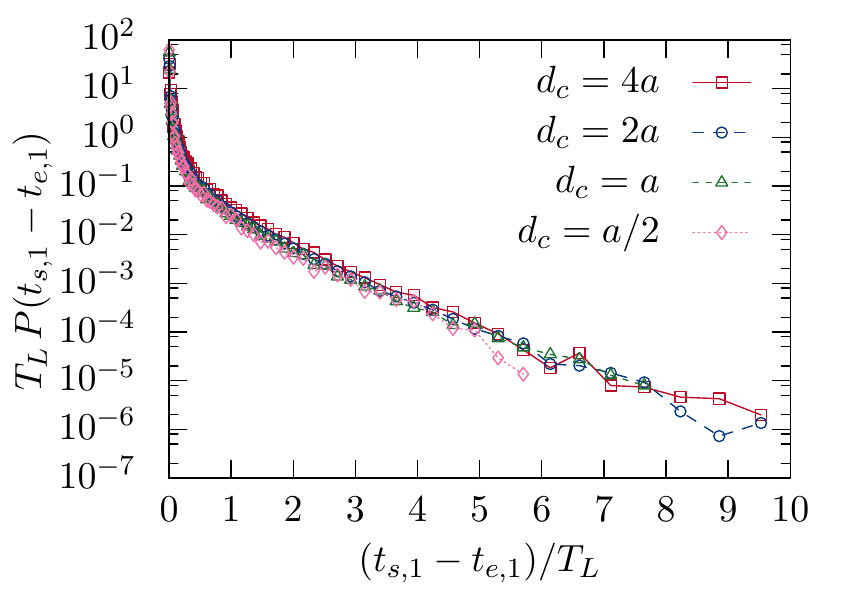}
\includeSubGraphics{(b)}{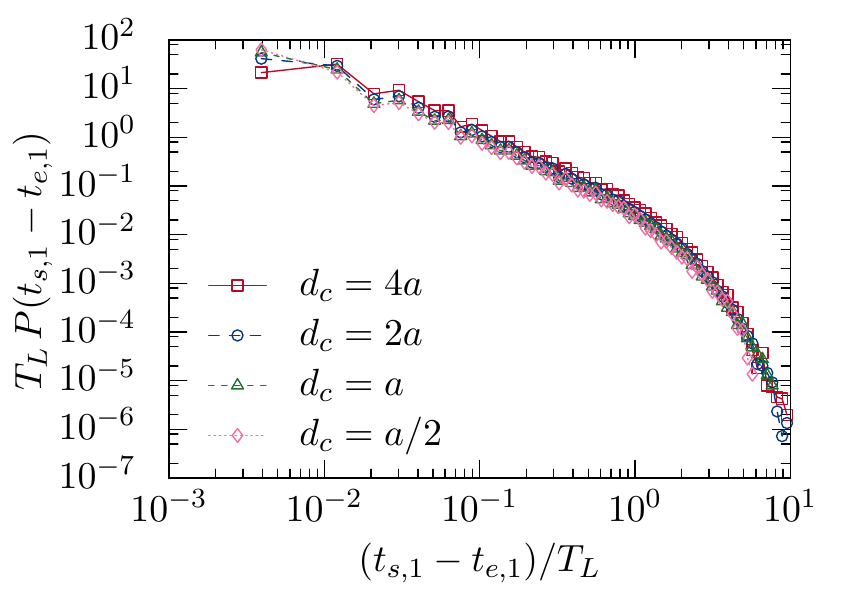}
\caption{(Color online)
Lin-log (panel a) and log-log (panel b)
representations of the distribution of
time spent together by pairs of particles during
their first encounter. Several values of the critical distance $d_c$
are shown; the similarity between the curves shows that the dependence
on $d_c$ is very weak (note that $d_c \ll \eta$ for all the values of $d_c$
shown on the figure).
The Stokes number used here is ${\rm St} = 1.5$.
}
\label{fig:Dt1_diff_dc}
\end{figure*}

Fig.~\ref{fig:Dt1_diff_dc} shows the distribution of the time particles
spend together during their first encounter $\Delta T_1 $. The data corresponds
to particles with a Stokes number ${\rm St} = 1.5$. Panel~(a)
shows the PDF of $\Delta T_1$ in lin-log scale. In this figure,
$\Delta T_1$ is expressed in units of the large eddy turnover time, $T_L$,
the correlation time of the flow. Fig.~\ref{fig:Dt1_diff_dc} shows an
exponential decay of the PDF, with a characteristic time of the order of the
large eddy turnover time. Thus, particles can spend together a time which
is comparable to the large eddy turnover time (at least for the value
of ${\rm R}_\lambda$ used in our simulations).
Fig.~\ref{fig:Dt1_diff_dc}(a) also indicates that the PDF has a very sharp
maximum around $\Delta T_1 \approx 0$.
In fact, Panel~(b) shows the PDF in log-log units, and suggests a power law distribution of $\Delta T_1$.
The exponent of the power law measured here is of the order $-1.5$. As
$\int_\epsilon x^{-1.5} \dd x$ diverges when $\epsilon \rightarrow 0$, the PDF
necessarily
saturates for time separations $\lesssim \Delta t$, where $\Delta t$ is the time at
which we saved trajectories.
Fig.~\ref{fig:Dt1_diff_dc} shows the PDF of $\Delta T_1$, determined with
several values of the critical distance $d_c$, very small compared to
the Kolmogorov scale $\eta$ (we have $a/\eta \approx 1/12$).
The PDFs are remarkably independent of $d_c$, at least provided
the ratio $d_c/\eta$ is small, where $\eta$ is the Kolmogorov length.

\begin{figure*}
\includeSubGraphics{(a)}{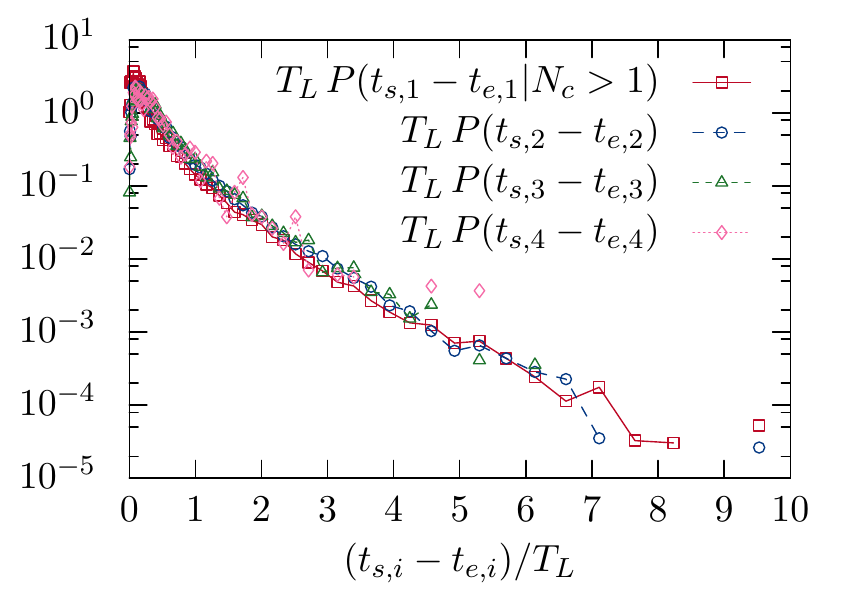}
\includeSubGraphics{(b)}{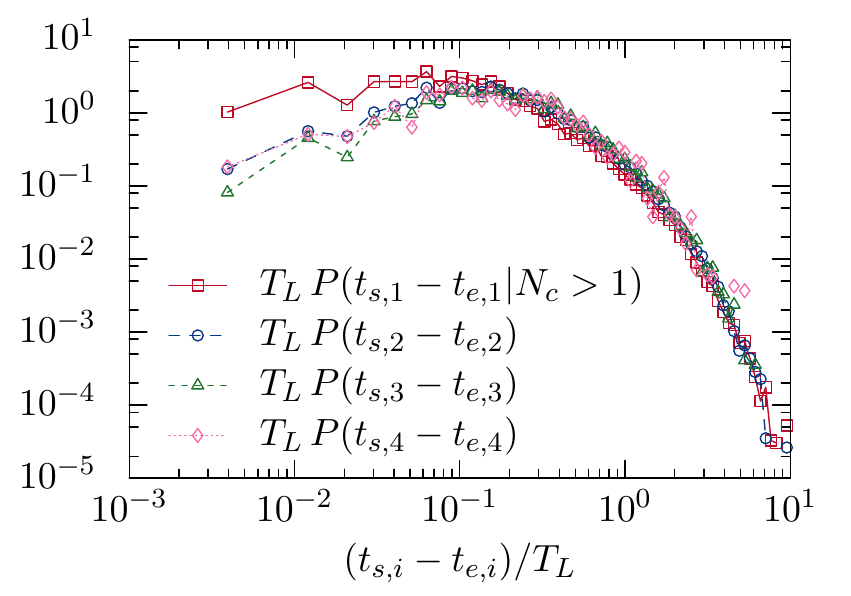}
\caption{(Color online)
Lin-log (panel a) and log-log (panel b)
representations of the distribution of
time spent together by pairs of particles during
second (circle), third (upward pointing triangle), fourth (diamond)
encounters. These distributions are remarkably similar.
For comparison, the distribution of time that particles spend together during
their first encounter, conditioned on the fact that they will meet again,
is shown (square symbols). The cuspy distribution, observed for
$t_{s,1} - t_{e,1}$, is not seen for these distributions.
The Stokes number used here is ${\rm St} = 1.5$.
}
\label{fig:Dtn_diff_dc}
\end{figure*}

Fig.~\ref{fig:Dtn_diff_dc} shows the probability distribution function
for $\Delta T_n$, with $n =2, 3, 4 $ ($n > 1$). The PDFs still exhibit for
very long values of $\Delta T_{n}$ exponentially decaying tails, see panel~(a), with the
same decay rate as obtained for $\Delta T_1$. However, the short time behavior
of the PDF does not exhibit the very large peak seen in Fig.~\ref{fig:Dt1_diff_dc}. In fact, the probability distribution does not
exhibit any power law at short values of $\Delta T_n$, as seen in panel~(b).

\begin{figure*}
\includeSubGraphics{(a)}{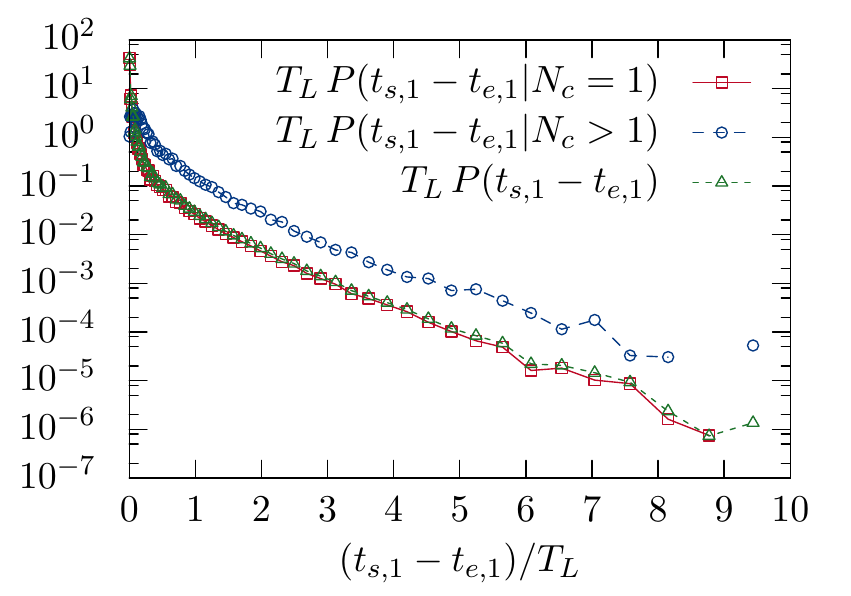}
\includeSubGraphics{(b)}{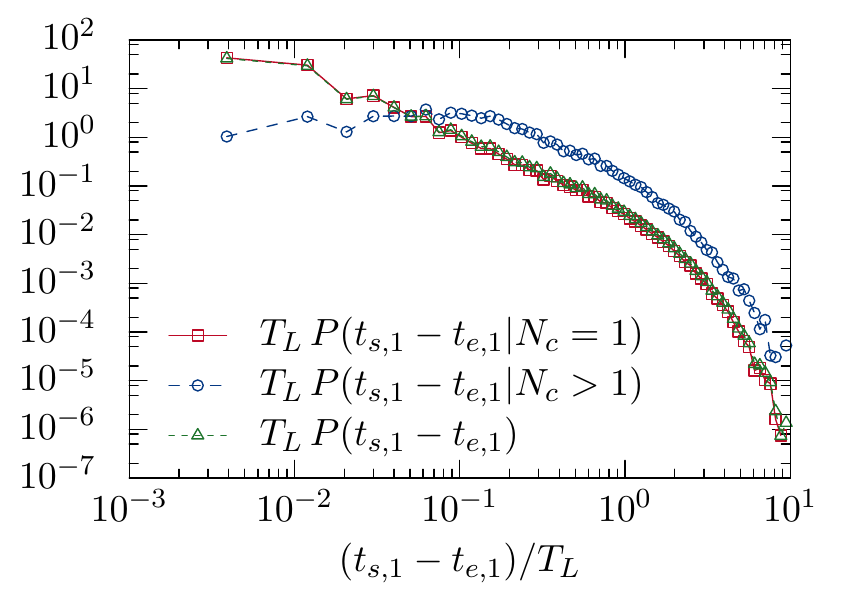}
\caption{(Color online)
The distribution of the time spent together by two particles
during their first encounter,
$p(t_{s,1} - t_{e,1})$ (upward pointing triangles), and the same distribution
conditioned on the fact that trajectories will separate after their first
encounter and not meet again ($N_{\rm c} = 1$; square symbols).
Finally the same PDF is shown again, this time conditioned on the fact that the particles
will meet again ($N_{\rm c} \ge 2$; circles).
The statistics is dominated by pairs that collide only once.
The large probability, hence the power-law distribution, that two particles
spend a short time together comes from trajectories with $N_{\rm c} = 1$.
Both conditional probabilities show the exponential tail for large times.
The Stokes number used here is ${\rm St} = 1.5$.
}
\label{fig:Dt1_decomposed}
\end{figure*}

The difference between the statistical properties of $\Delta T_1$ and
$\Delta T_n$ for $n \ge 2$ is therefore restricted to the short time behavior.
In physical terms, the probability that the two particles do not spend much
time together is much larger during the first encounter than during the
following one. Particles can spend a short time when they are impacting each
other with a large velocity difference. One may surmise that in such a case,
particles will separate very fast, and not get into close contact afterward.
In other words, the events leading to several contacts are unlikely to have
initially a small value of $\Delta T_1$. To actually check this,
Fig.~\ref{fig:Dtn_diff_dc} also shows the PDF of $\Delta T_1$, conditioned on
the fact that the two trajectories will come into contact more than one
time, see the curve with the square symbols. As expected, conditioning the
probability of $\Delta T_1 $ on the fact that there are more than one encounter
between the trajectories significantly reduces the probability for the particles
of separating very fast. In this spirit, Fig.~\ref{fig:Dt1_decomposed} shows
the PDF of $\Delta T_1$ conditioned on having several successive encounters
between the trajectories ($N > 1$), or simply one ($N = 1$), together with the
total PDF of $\Delta T_1$. As the probability of having multiple encounters
between the two particles remains relatively small, the PDF of $\Delta T_1$
is extremely close to the PDF of $\Delta T_1$ conditioned on having $N = 1$.

\begin{figure*}
\includeSubGraphics{(a)}{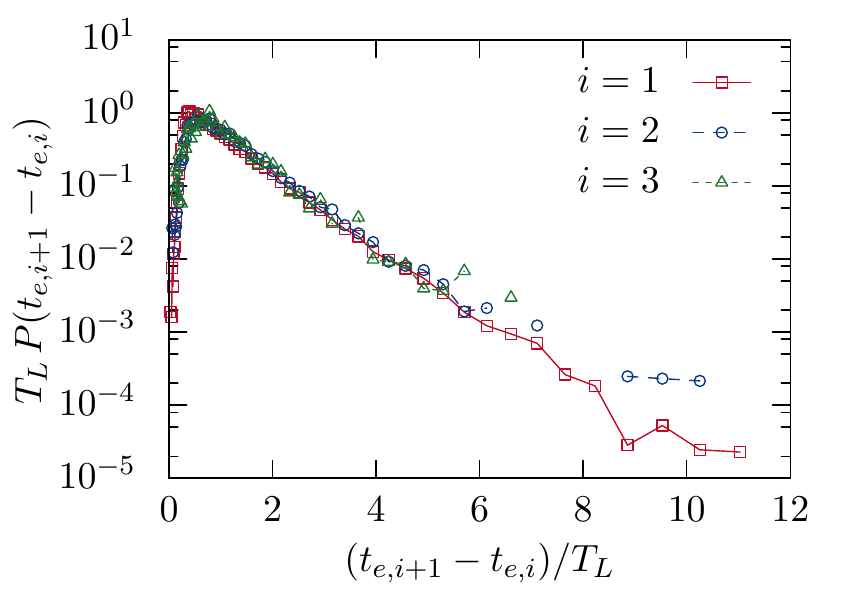}
\includeSubGraphics{(b)}{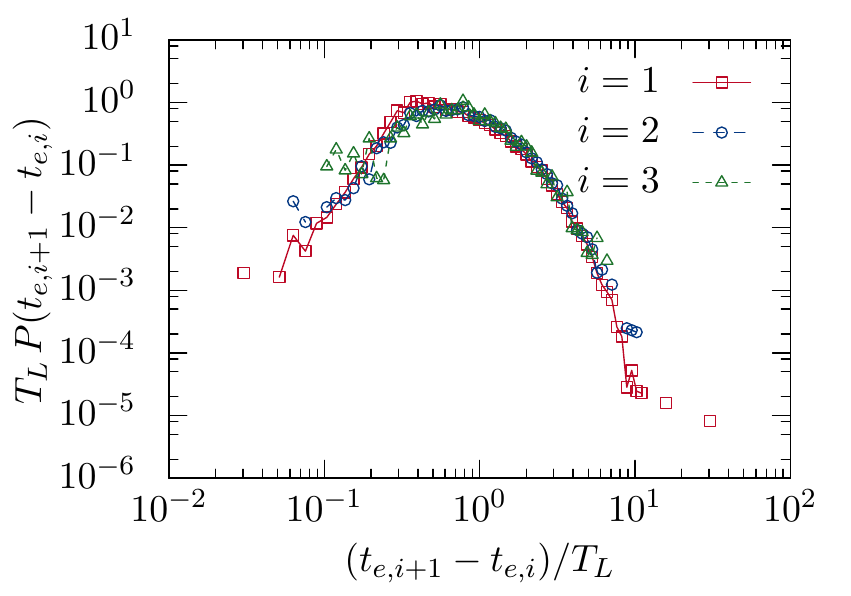}
\caption{(Color online)
Statistics for successive encounters between two particle trajectories.
The PDF of $t_{e,i+1} - t_{e,i}$ in lin-log scales, $t_{e,i}$ being
defined in Fig.~\ref{fig:times_illustration}.
Panel (a) shows the data in lin-log scaling, while panel (b) shows the same data in log-log scaling.
The PDFs do not depend on $i$, suggesting a self similar process.
The Stokes number used here is ${\rm St} = 1.5$.
}
\label{fig:Dt_s_n}
\end{figure*}

Aside from the difference between the PDF of $\Delta T_1 $ and $\Delta T_n$
at short times, it appears that the process leading to subsequent encounters
between particle trajectories is largely self-similar {\it i.e.} does not
depend much on the index $n > 1$. This effect is strengthened by studying the
difference between the time it takes for two trajectories to come into
contact again. To this end, Fig.~\ref{fig:Dt_s_n} shows the probability
distribution function of $\Delta T^e_n = (t_{e,n+1} - t_{e,n})$. The
PDF is found to be remarkably similar (independent of $n$). As it was the
case for $\Delta T_n$, the PDF has an exponential tail at large values of
$\Delta T^e_n$. The distribution however peaks at a finite value of
$\Delta T^e_n \approx 0.6 T_L$.

\subsection{Stokes number dependence  of the contact-time statistics}
\label{subsec:Stokes_stat_traj}

\begin{figure*}
\includeSubGraphics{(a)}{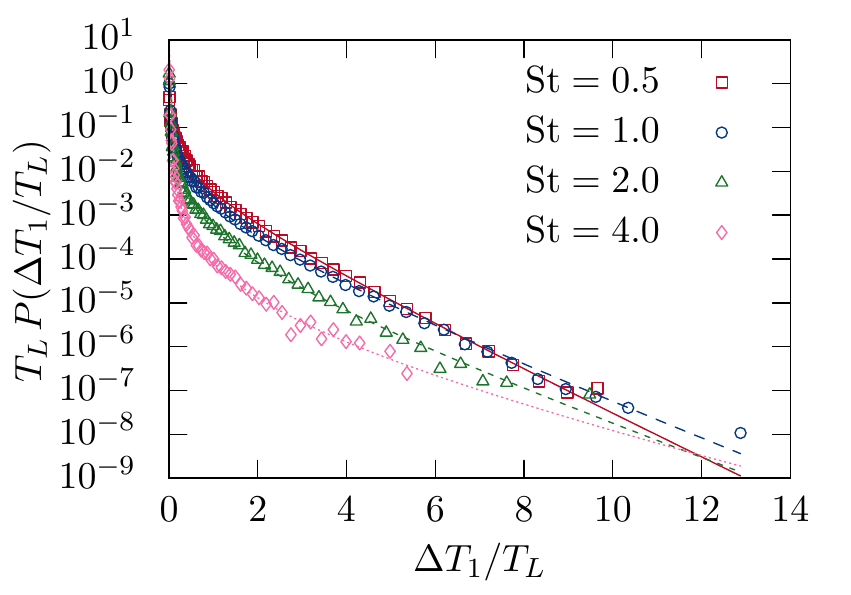}
\includeSubGraphics{(b)}{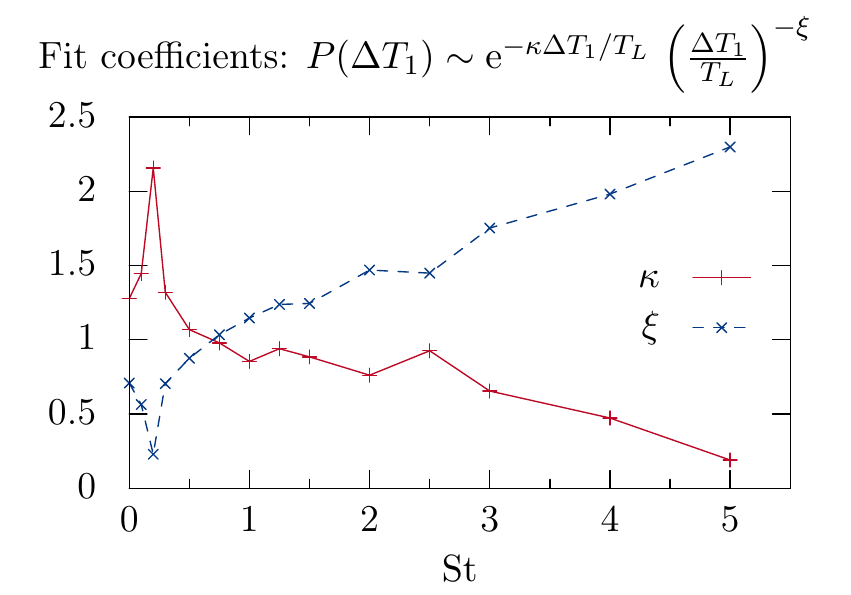}
\caption{(Color online)
Fit of the PDF of $\Delta T$ by \eqref{eq:fit_PDF} at different values
of the Stokes number.
Panel (a) shows the quality of the fit, for four values of
the Stokes number, indicated in the legend.
The quality of the fit degrades at small values
of ${\rm St}$, suggesting that a more complicated functional form
may be needed.
Panel (b) shows the variation of the exponent $\xi$ and of the
coefficient $\kappa$ determined by fitting the PDF, as a function
of ${\rm St}$.
}
\label{fig:Stokes_dependence}
\end{figure*}

The general picture, shown in subsections~\ref{subsec:obs}
and \ref{subsec:Stat_traj} for particles with a Stokes number
${\rm St} = 1.5$, has been found to be qualitatively unchanged
when the Stokes number is varied. However, the details differ quantitatively.
This can be seen by representing the distribution of the time difference
$\Delta T_1$ by an asymptotic fit of the form:
\begin{equation}
P(\Delta T_1) \approx {\cal N} (\Delta T_1/T_L)^{-\xi} \exp( - \kappa  \Delta T_1 /T_L)
\label{eq:fit_PDF}
\end{equation}
The coefficient $\cal{N}$ in \eqref{eq:fit_PDF} is simply adjusted to enforce
that the PDF is properly normalized. The coefficients $\xi$ and $\kappa$
are determined by fitting the PDFs. The quality of the fit is very
good, as shown in Fig.~\ref{fig:Stokes_dependence}~(a),
at least for
$0.3 \lesssim {\rm St}$. The fitting parameters are found to depend
very significantly
on ${\rm St}$, see Fig.~\ref{fig:Stokes_dependence}~(b). In the very small
Stokes number limit, the exponent $\xi$ diminishes, suggesting that the
distribution $P(\Delta T_1) $ is becoming closer to a purely exponential
distribution. In the opposite limit, the coefficient $\kappa$ decreases,
whereas the power $\xi$ seems to increase.

\section{Velocity difference between colliding particles }
\label{sec:vel_diff}

The time that two ghost particles spend in \lq contact' (that is the length
of time over which the separation of their centers satisfies $d\le 2a$)
is determined by two factors: the impact parameter,
and the relative velocity of the collision. The data in
section~\ref{sec:contactstats} show two striking aspects of the
contact-time statistics. Firstly, the statistics of the first contact are
different from those of all the others. Secondly, there is evidence for
a power-law regime in the distribution of the first contact time. In this
section we discuss how both of these
observations can be explained in terms of properties of the distribution of
collision velocities.

\subsection{Multiple collisions happen at slow relative velocities }
\label{subsec:spurious_vel}

\begin{figure*}
\includeSubGraphics{(a)}{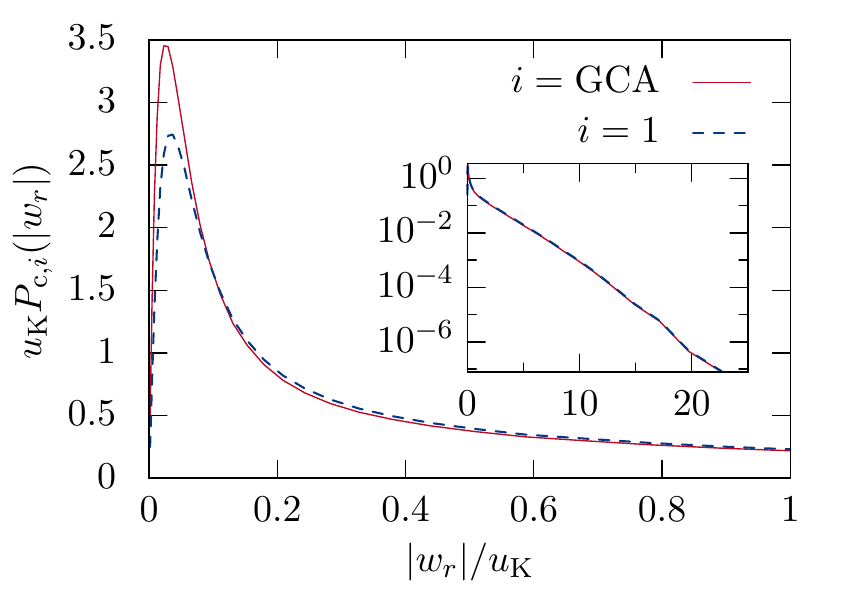}
\includeSubGraphics{(b)}{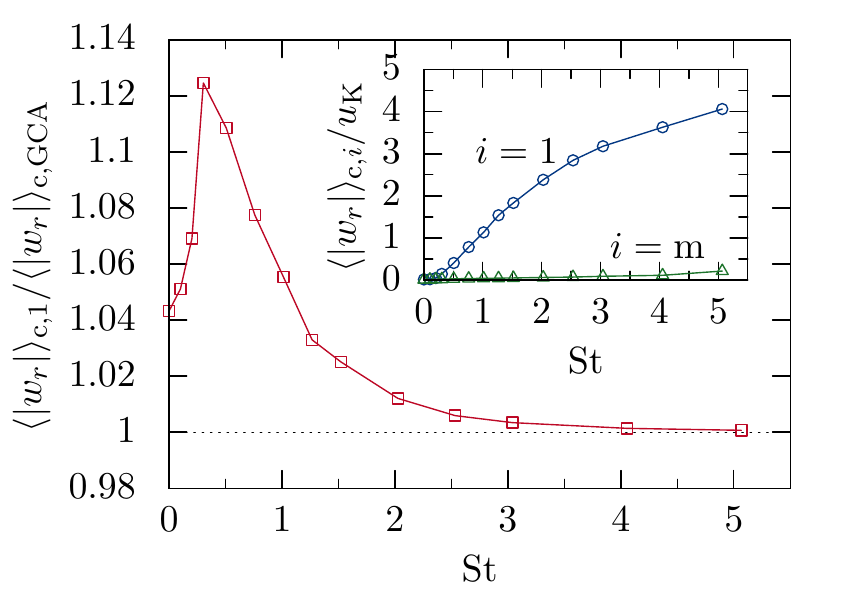}
\caption{(Color online)
Probability of the radial relative collision velocity.
In panel (a) the full line corresponds to the PDF as obtained when taking into account all collisions, as in the GCA.
The dashed line gives the PDF of solely first collisions.
The figure has been obtained with a value of
the Stokes number ${\rm St} = 1.0$. The velocities are expressed
in terms of
the velocity $u_{\rm K} = (\eta_{\rm K} / \tau_{\rm K})$ at the smallest scale.
Panel (b) shows the ratio of the two mean radial relative collision velocities $\langle \lvert w_r \rvert\rangle_{\mathrm{c},1}$ and $\langle \lvert w_r \rvert\rangle_{\mathrm{c},\mathrm{GCA}}$ for different Stokes numbers.
The former takes into account only first collisions, the latter incorporates all collisions detected within the framework of the GCA.
Furthermore the inset in panel (b) shows $\langle \lvert w_r \rvert\rangle_{\mathrm{c},1}$ and the radial relative velocity of multiple collisions $\langle \lvert w_r \rvert\rangle_{\mathrm{c},\mathrm{m}}$ in terms of $u_\mathrm{K}$.
}
\label{fig:gamma_sp_conditioned}
\end{figure*}

The probability of the radial relative velocity, defined as $w_r = \delta \vec{v} \cdot \delta \vec{r} / \lvert \delta \vec{r} \rvert$, is shown in Fig.\,\ref{fig:gamma_sp_conditioned} for colliding particles and at ${\rm St} = 1$.
We remind that the PDF for colliding particles is different from the one for all particles in contact \cite{wang00jfm}.
In panel (a) this PDF is shown for two different situations.
In the first one, all colliding particles are taken into account,
as is the case in the GCA.
In the other one, only pairs that collide for the first time are taken into account.
In both cases the bulk of the PDF is located at small values of
$\lvert w_r \rvert$, and exhibits an exponential tail at very large
collision velocities (see inset).
A close comparison of the two probabilities shows that the
contribution of small
relative velocities corresponding to first collisions is smaller than for
all particle pairs detected with the GCA.
This suggests that the error in the GCA stems mainly
from collisions with small relative velocities.
The right panel of Fig.\,\ref{fig:gamma_sp_conditioned} extends this
observation to different Stokes numbers by comparing the mean radial relative
collision velocities obtained with both schemes.
For Stokes numbers $\lesssim 2$ the collisions corresponding to
particles in contact for the first time
show on average slightly larger radial relative velocities.
However, the averaged relative velocities for particle pairs colliding
more than one time are significantly smaller than both, the velocities of first collisions and the velocities in the GCA.
This can be seen in the inset of the right panel.
The observations summarized in Fig.~\ref{fig:gamma_sp_conditioned} therefore
demonstrate that, multiple collisions occur at small relative velocity.

It has been argued that the collision rate between particles in a turbulent
suspension can be resolved into two components, as represented by equation
(\ref{eq:advcaust}) \cite{falkovich02nat,wilkinson06prl}.
In the decomposition \eqref{eq:advcaust}, the term
$\Gamma_{\rm adv}$ represents collisions
due to shearing motion, which occur with relative velocities of order
$a/\tau_{\rm K}$, whereas the term $\Gamma_{\rm caust}$ represents
collisions between particles which are moving relative to the flow,
and on different branches of a phase-space manifold,
separated by a caustic. The relative velocity of the collisions which
contribute to $\Gamma_{\rm caust}$ is much higher, of order
$(\eta/\tau_{\rm K})f({\rm St})$, where
$f({\rm St})$ is an increasing function, which is of order unity at ${\rm St}=1$.
The ideas underlying
this decomposition also explain why the statistics of the first
contact time are
different from those of all the subsequent contacts. According to this picture, multiple
collisions are almost exclusively due to the advective collision mechanism, and are very
unlikely for caustic-induced collisions because the high relative velocity
rapidly moves the particles out of proximity. In the advective process multiple collisions arise
because of temporal fluctuations of the shear rate tensor \cite{gustavsson08njp}.
This model explains why the first collision has different contact time statistics, and why the
multiple collisions have small relative velocities.

\subsection{Power-law distribution of contact times }
\label{subsec:power-law}

\begin{figure}
\includegraphics{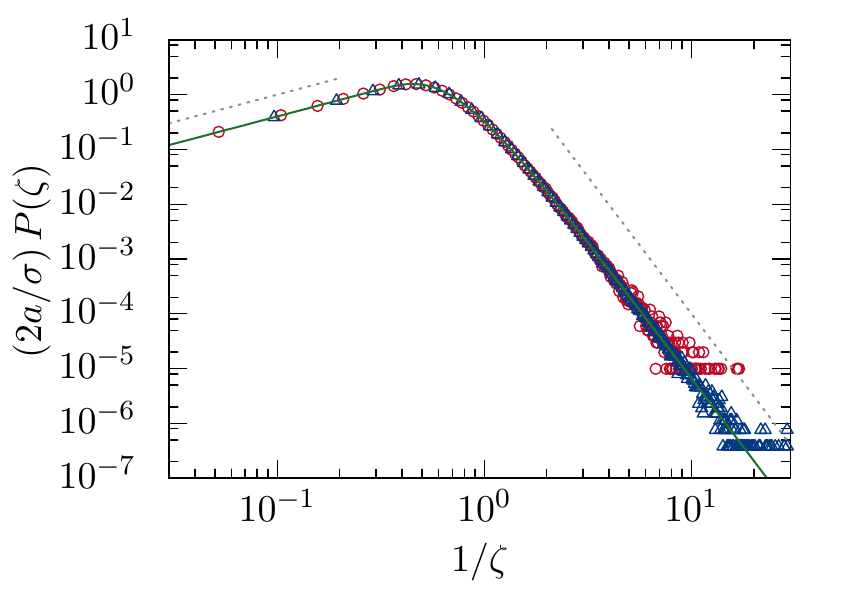}
 \caption{(Color online)
Comparison between the calculation of the time $\Delta T$
spent by two
particles together, by simulating directly a gas of Maxwell particles (symbols),
and the result of the analytic formula \eqref{eq:max_gas}
(continuous line).
The different symbols correspond to simulations with larger (circles) and smaller (triangles) particles.
The time $\Delta T$
is expressed in terms of $\zeta$, defined by \eqref{eq:zeta}.
The behavior of the distribution decays for large values of $\Delta T$
as $\sim \Delta T^{-5}$.
This limit, as well as the short time behavior $\sim \Delta T$ are shown as dotted lines.
\label{fig:maxwell}
 }
\end{figure}

The evidence
for a power-law in the distribution of the contact time for the
first collision is one of the conspicuous results
of Sec.~\ref{sec:contactstats}.
In order to understand the origin of such power-laws, we consider
the following simplified model. We computed directly
the distribution of the time $\Delta T$ that two particles in a gas of
particles, with a Maxwellian distribution of velocity spend within a distance
$2a$ from each other.
The root mean square of one velocity component of these particles is taken to be $\sigma$.
This model corresponds to the very large ${\rm St}$ limit of inertial particles in a turbulent flow.
In that case,
the gas of particles
reduces to particles each moving with its own velocity, distributed
according to the Maxwell distribution~\cite{Abr75}.
A simple calculation leads to the following PDF
of the time that two particles spend together:
\begin{align}
P(\Delta T ) &= 2 \frac{\sigma}{a} \frac{1}{\zeta} \Bigl( 1 - \exp( - \zeta^2 ) \bigl[ 1 + \zeta^2 + \frac{1}{2} \zeta^4 \bigr] \Bigr)
\label{eq:max_gas} \\
\text{with}\qquad \zeta & = \frac{2 a}{\sigma\,\Delta T} \label{eq:zeta}
\end{align}
In the limit of long times, $ \Delta T \rightarrow \infty$,
$\zeta \rightarrow 0$, the PDF in Eq.\,\eqref{eq:max_gas}
reduces to:
\begin{equation}
P(\Delta T) \propto \frac{1}{\Delta T^5}
\label{eq:pdf_gaus}
\end{equation}
This result was checked directly by generating a gas of Maxwellian
particles and determining the time particles spend together (Fig.\,\ref{fig:maxwell}). The distribution
of $\Delta T$ is found in excellent agreement with \eqref{eq:max_gas}.
We furthermore note, that Equation~\eqref{eq:max_gas} displays another power law for short contact times.
But this behavior is only apparent for $\Delta T < 2a / \sigma \ll\tau_{\rm K}$.
In our DNS of the turbulent transport of particles we do not resolve these time scales.

Let us consider the implications of the model leading to
equation (\ref{eq:pdf_gaus}) for the distribution of first contact times.
The important point is to observe that a power-law in the distribution
of \emph{small} relative velocities leads to a power-law distribution of
\emph{long} contact times. The data in figure \ref{fig:Dt1_diff_dc} show
a power-law distribution in the contact times at \emph{short} times.
We should, however, remember that the first contacts are a combination
of caustic-mediated collisions (with high relative velocity) and advective collisions
(with low relative velocity). We propose that the power-law distribution
of the contact time results from the low-velocity tail of the caustic-mediated
collisions having a PDF which can be approximated by a power-law at small
velocities.

\section{Conclusion}
\label{sec:concl}

We have studied the collision rate in turbulent suspensions,
with the aim of evaluating the systematic errors made while using
the GCA for particles which aggregate upon collision.
To this end, we have compared the results using the GCA, and a
more realistic algorithm, which consists in replacing one of the particles
that underwent collision by a particle from a \lq reservoir'.
The error we find by using the GCA is as large as $\sim 15\,\%$ at very small
Stokes number, and tends to decrease when ${\rm St}$ increases.

We investigated the statistics of the multiple collisions which are
the source of the spurious collisions in the GCA. We found that,
after the first collision, multiple collisions  involving the
same pair of particles occur with a probability that decays exponentially
with the number of collisions, $N_{\rm c}$. We also studied the time
that particle trajectories spend close to one another. We find the PDF of the
time spent by two particles within a distance smaller than a critical value $d_c$
exhibits an exponential tail at very long times. At shorter times the contact
time PDF of the first collision obeys a power law.

\begin{acknowledgments}
We thank Bernhard Mehlig and Kristian Gustavsson for
discussions, and Matth\"aus B\"abler for pointing out \cite{Ott:05}.
AP and EL acknowledge the financial support of ANR (contract TEC 2). The numerical calculations have been performed at the PSMN computer center of the
Ecole Normale Sup\'erieure in Lyon.
MW and AP were supported by the EU COST action MP0806 \lq Particles in Turbulence'.

\vfill

\end{acknowledgments}

\appendix

\bibliographystyle{apsrev4-1}
\bibliography{references}

\appendix

\section{}

In this appendix we consider the equivalence of two
different approaches to determining the true rate of first contact
collisions, $\Gamma_1$, and therefore provide evidence that
$\Gamma_1$ is indeed the physically appropriate definition of the
collision rate, for a class of systems where particles react only at
first contact.
One approach is to use ghost particles
and count the rate of multiple collisions, so that
$\Gamma_1=\Gamma_{\rm GCA}-\Gamma_{\rm m}$. The alternative
approach is to use one of the two \lq substitution schemes',
described in subsection \ref{sec:detect}, leading to rates
$\Gamma_{{\rm Re}1}$ and $\Gamma_{{\rm Re}2}$. In this appendix
we show that these estimates differ as a result of finite
particle density effects. We provide evidence however that they
become equal in the limit of very dilute systems (that is, as
the particle density approaches zero).

\begin{figure*}
\includeSubGraphics{(a)}{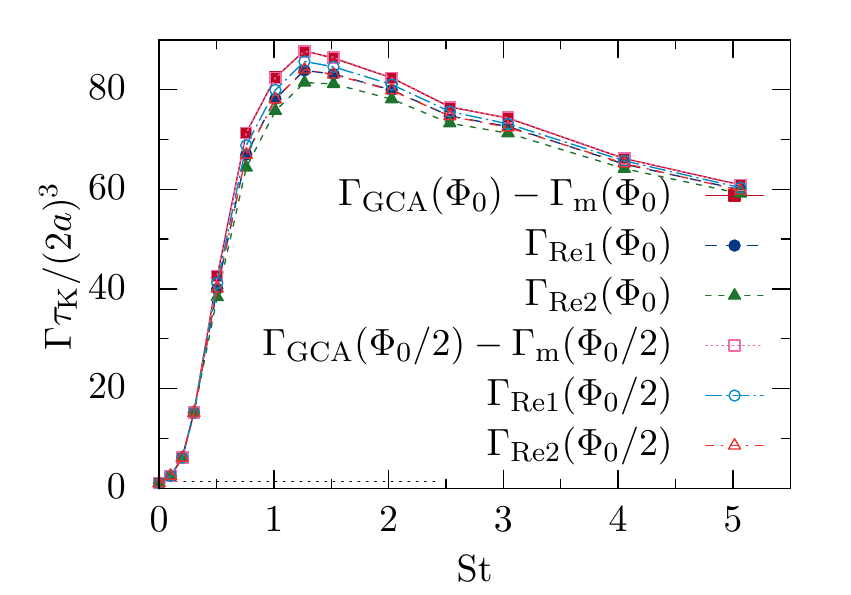}
\includeSubGraphics{(b)}{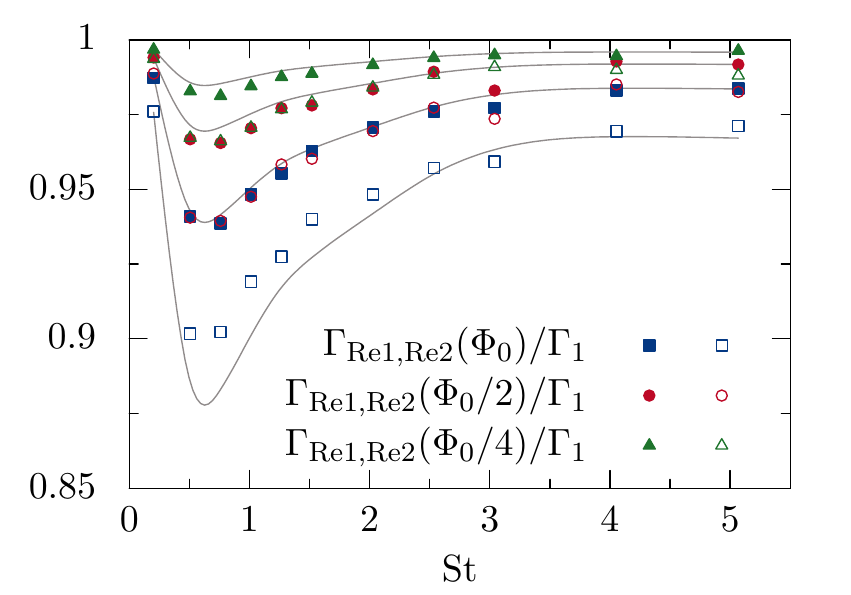}
 \caption{(Color online)
Comparison between $\Gamma_1=\Gamma_{\rm GCA}-\Gamma_{\rm m}$, and
the collision rates obtained by using the substitution algorithms
($\Gamma_{{\rm Re}1}$ and $\Gamma_{{\rm Re}2}$).
The left panel shows the raw data. The right panel
shows $\Gamma_{{\rm Re}i}/\Gamma_1$ as
a function of the Stokes number, for those runs listed in Table~\ref{tabl:1}
for which $\Phi_0 = 4.5 ~ 10^{-5}$. The open (filled) symbols correspond to
the algorithm {Re2}, ({Re1}). The full lines are deduced from each other
by multiplication by a factor $2$, and suggest that the dependence of the
difference $\Gamma_1 - \Gamma_{{\rm Re}i}$ behaves linearly with the density
of particles in the system.
}

\label{fig:corr_Gamma}

\end{figure*}

Fig.~\ref{fig:corr_Gamma} shows the collision rates produced by
these algorithms, as well as $\Gamma_1$, obtained at
several values of the volume fraction $\Phi$: $\Phi_0$, $\Phi_0/2$, and $\Phi_0/4$.
The numerical results show that

\begin{enumerate}
\item[(1)]
$\Gamma_{{\rm Re}2}(\Phi) < \Gamma_{{\rm Re}1}(\Phi) < \Gamma_1= \Gamma_{\rm GCA} - \Gamma_{\rm m}$

\item[(2)]

$\Gamma_{{\rm Re}2}(\Phi/2) = \Gamma_{{\rm Re}1}(\Phi)$

\end{enumerate}

A more careful analysis of the data reveals that when $\Phi \rightarrow 0$,
both $\Gamma_{{\rm Re}1}$ and $\Gamma_{{\rm Re}2}$ tend to $\Gamma_{\rm GCA} - \Gamma_{\rm m}$,
with corrections which are linear in $\Phi$.
Thus, our numerical results demonstrate that the results from the three
different algorithms agree in the dilute limit, $\Phi \rightarrow 0$.

The numerical results can be understood in the following way. We introduce
the coefficients $R = \Gamma n$, which determine the rate of collision of a
test particle moving in a medium containing other particles with number density $n$.
The ergodic assumption
implies that the actual collision rate can be inferred from the long-time
behavior of a single test trajectory.

The quantity $n\Gamma_{\rm GCA}$ is just the total rate of collision along the test trajectory,
and $n\Gamma_{\rm m}$ is the rate for collisions in which the test particle encounters
the same target particle more than once. For systems where there is coalescence
on contact, only the rate of first collision is of interest.
The preferential concentration effect enhances the rate of subsequent collisions with
further
particles, once a pair of particles has collided. However, in the limit
$n \rightarrow 0$, the time between collision events approaches infinity
($\propto 1/n$). For this reason, in the limit as $n \rightarrow 0$ the first
contact collision rate is precisely $n\Gamma_{{\rm Re}i}$.

In terms of the long-time average over trajectories, the algorithm
which yields $\Gamma_{{\rm Re}2}$ changes
the position of the trajectory, therefore destroying the possibility of further
collisions with the surrounding environment. In addition, a different
realization of the surrounding background particles is chosen.
In contrast, for the algorithm which yields $\Gamma_{{\rm Re}1}$,
either the position of the test trajectory, or the
background of surrounding particles is changed, each with a probability $1/2$.

Changing the position of the test particles, or the surrounding background
results first of all in preventing any further collision between
the pair of particles that came into contact. The simplest
approximation for the probability per unit time of a collision between the
test particle and a third particle, after a contact has been detected, is
proportional to $R$. This suggests that the approximations $\Gamma_{{\rm Re}1}$
or $\Gamma_{{\rm Re}2}$ miss
a correction of order ${\cal O}(R \tau)$, where $\tau$ is a characteristic
time over which the surrounding particles rearrange.
Moreover, the algorithm yielding $\Gamma_{{\rm Re}2}$
misses twice as many collisions with a third
particle as $\Gamma_{{\rm Re}1}$, which is also consistent
with our numerical observations.
These arguments suggest that
\begin{equation}
\label{eq:A1}
\Gamma_{{\rm Re}i} = \Gamma_1 [1 - O(i R \tau)]\ .
\end{equation}
As $R \propto n$, this expression justifies the numerical observation that
$\Gamma_{{\rm Re}1,2}$ are smaller than $\Gamma_{1}$, and differ from it by a
quantity proportional to $n$.

\end{document}